\documentclass{JAC2000}
\usepackage{graphicx}
\begin{document}
\title{PULSED SUPERCONDUCTIVITY ACCELERATION}
\author{M. Liepe\thanks{liepe@sun52a.desy.de} for the TESLA Collaboration\\ Deutsches Elektronen-Synchrotron DESY, D-22603 Hamburg, Germany}
\maketitle
\begin{abstract} 
The design of the proposed linear collider TESLA is based on 9-cell 1.3 GHz superconducting niobium cavities, operated in pulsed mode \cite{TESLACDR}. Within the framework of an international collaboration the TESLA Test Facility (TTF) has been set up at DESY, providing the infrastructure for cavity R\&D towards higher gradients \cite{TTFDR}. More than 60 nine-cell cavities were tested, accelerating gradients as high as 30 MV/m were measured. In the second production of TTF-cavities the average gradient was measured to be 24.7 MV/m. Two modules, each containing eight resonators, are presently used in the TTF-linac. These cavities are operated in pulsed mode: 0.8 ms constant gradient with up to 10 Hz repetitions rate. 
We will focus on two aspects: Firstly, the cavity fabrication and treatment is discussed, allowing to reach high gradients. Latest results of single cell cavities will be shown, going beyond 40 MV/m. Secondly, the pulsed mode operation of superconducting cavities is reviewed. This includes Lorentz force detuning, mechanical vibrations (microphonics) and rf field control.
Both topics meet the upcoming interest in superconducting proton linacs, like the sc driver linac for SNS (Spallation Neutron Source).
\end{abstract}
\section{INTRODUCTION}
The aim of the  cavity program - launched by the TESLA collaboration in 1992 - is to explore the technology of high gradient superconducting cavities and to demonstrate the performance and reliability of a superconducting linac operated in pulsed mode. The infrastructure for the treatment, assembly and test of superconducting cavities has been established at DESY. A full integrated system test with beam is done at the TTF-linac \cite{TTFDR}. In a first step the design goal for the cavities of the TTF-linac was set to the value $E_{acc}=$15 MV/m.  The aim was to gradually approach the accelerating gradient of 22 MV/m required for TESLA with a centre-of-mass energy $E_{cm}$ = 500 GeV \cite{Brinkmann}.
\section{PERFORMANCE OF THE SUPERCONDUCTING TTF-CAVITIES}
\begin{figure}[htb]
\centering
\includegraphics*[width=65mm]{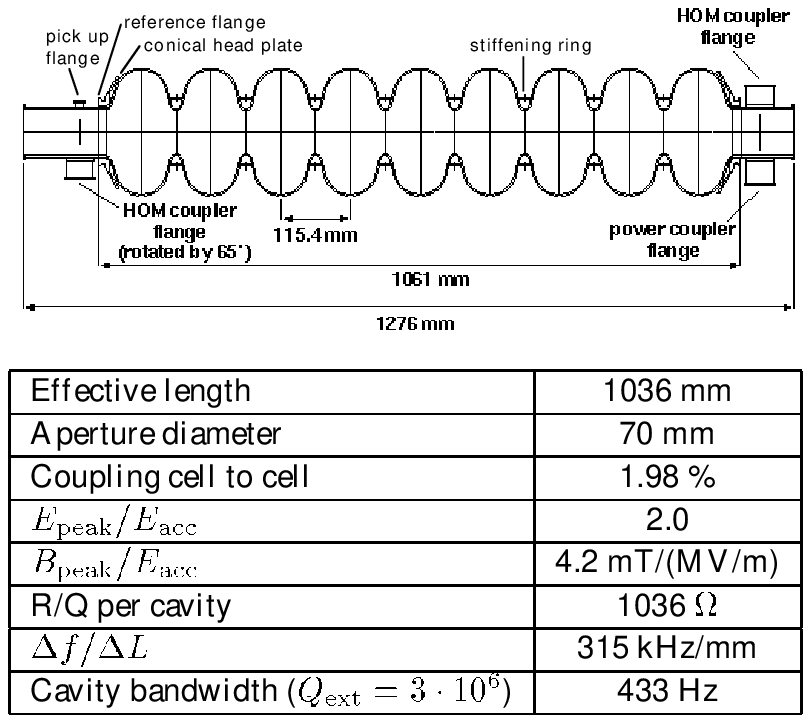}
\caption{Cross section and selected design parameters of the 1.3 GHz TTF 9-cell cavity.}
\label{cavity}
\end{figure}

The TTF 9-cell cavities are fabricated from $RRR=$ 300 niobium by electron-beam welding of half cells that are deep-drawn from niobium sheet metal, see Fig.~\ref{cavity}.
The cavities have been ordered at four European companies and have been prepared and tested at DESY \cite{cavpaper}. The presently used standard cavity preparation before the vertical test consists of the following steps:
\newline
\par\textbullet \ removal of the damage layer by buffered chemical \par \ \ \ polishing (BCP): 80 $\mu$m  from the inner  and 30  $\mu$m \par \ \ \ from the outer cavity surface
\par\textbullet \ 2 hours heat treatment at 800 $^{\circ}$C
\par\textbullet \ 4 hours heat treatment at 1400 $^{\circ}$C with titanium getter
\par\textbullet \ removal of the titanium layer by 80 $\mu$m inner \par \ \ \ and  30  $\mu$m outer BCP
\par\textbullet \ field flatness tuning 
\par\textbullet \ final 20 $\mu$m removal from the inner surface by BCP
\par\textbullet \ high pressure rinsing (HPR) with ultrapure water
\par\textbullet \ drying by laminar flow in a class 10 cleanroom
\par\textbullet \ assembly of all flanges, leak-check
\par\textbullet \ 2 times HPR, drying by laminar flow and assembly \par \ \ \ of the input antenna with high external $Q$.
\subsection{Vertical Test Results}
\begin{figure}[htb]
\centering
\includegraphics*[height=75mm, angle=-90]{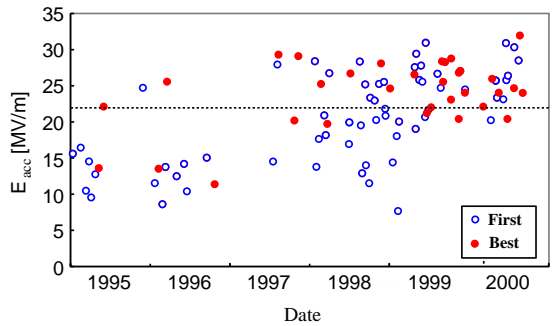}
\caption{Time development of the maximum gradient achieved in TTF 9-cell cavities. Shown is the cavity performance in the first test and in the best test. Note that several cavities have been tested only once (results labeled here as first test).}
\label{firstbest}
\end{figure}

\begin{figure}[htb]
\centering
\includegraphics*[height=70mm, angle=-90]{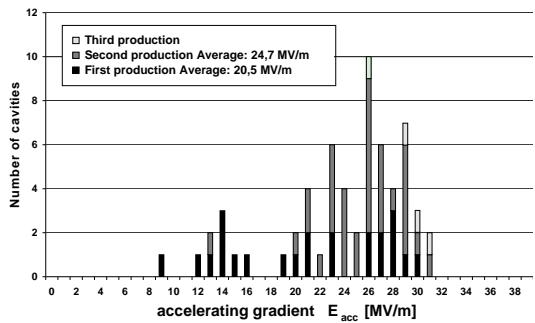}
\caption{Distribution of maximum gradient for the TTF 9-cell cavities (best test). }
\label{hist9cell}
\end{figure}

Up to date more than 60 nine-cell cavities have been tested  with cw rf-excitation in the vertical cryostat.  The time development of the test results is  summarized in  Fig.~\ref{firstbest}.  
The performance improvement over the past years is clearly visible. In particular, during the last year nearly all of the cavities performed above 22 MV/m. More important, the majority of these cavities reached their good performance already in the first test.
The distribution of the maximum gradient is shown in Fig.~\ref{hist9cell} for all 9-cell cavities. In the cavities with  low performance defects in the welds or in the bulk niobium were found \cite{cavpaper}. For the second cavity  production the welding technique was improved and all niobium sheets were eddy current scanned.  These cavities  reached an average accelerating gradient of 24.7 MV/m, thereby exceeding the required gradient for TESLA500 on average. All cavities of the third production which have been tested up to date showed a performance clearly above the TESLA specification.   
\subsection{Horizontal Test Results}
\begin{figure}[htb]
\centering
\includegraphics*[width=60mm]{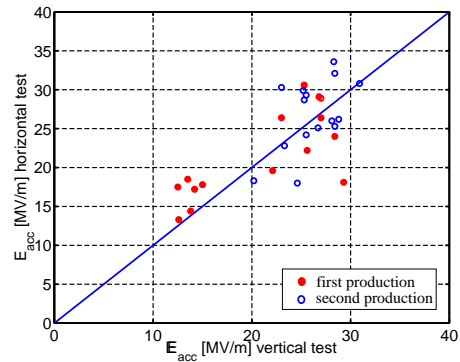}
\caption{Comparison of vertical and horizontal test results of cavities from the 
frist and the second production. The average accelerating gradient in the vertical tests with cw excitation is 23.6 MV/m, in the horizontal test with pulsed operation 24.2 MV/m.}
\label{vert_hor}
\end{figure}

After the cavities have passed the vertical acceptance test successfully, the helium vessel is welded to the head plates of the cavity, see Fig.~\ref{cavity}. A 20 $\mu$m removal from the inner surface follows. In the last preparation step before the horizontal test, the main power coupler is assembled to the high pressure rinsed cavity. The external $Q$ of the power coupler is typically $2\cdot10^6$. More than 30 cavites have been tested in pulsed mode operation (see Fig.~\ref{pulse}) in the horizontal cryostat (''CHECHIA''). The average gradient achieved in the vertical and the horizontal tests are quite similar as shown in Fig.~\ref{vert_hor}.  
In a few cases the performance was reduced in the horizontal test due to field emission. In other cavities the maximum gradient was improved by the fact that the cavites are operated in pulsed mode instead of the cw operation in the vertical test. These results demonstrate that the good performance of a cavity can be preserved after the accembly of the helium vessel and the  power coupler. 
\subsection{Cavity Performance in the TTF-Modules}
Eight 9-cell cavities, each with its own helium vessel and main power coupler, are assebled in a string in the cleanroom. The cavity string is than installed in a cryo-module. Up to date four accelerator modules have been built, three of them have been tested in the TTF linac (the third one replaced module \#1).  The cavities in the modules are operated in pulsed mode: 0.8 ms constant gradient with up to 10 Hz repetitions rate (see Fig.~\ref{pulse}).  Due to the fact that all cavities of a module are operated at the same gradient, the performance of a module is limitted by the cavity with the lowest performance. The measured average gradient was 16 MV/m, 20 MV/m and 22.5 MV/m for the modules \#1, 2 and 3, respectively. The module \#4 has been equipped with cavities, that have shown a performance of $E_{acc}\geq$25 MV/m in the vertical test. 
\section{CAVITY R\&D}
The single-cell cavity R\&D program focuses on possible simplifications in the cavity production and treatment and on pushing the cavity performance towards higher gradients. For a possible energy upgrade of TESLA to $E_{cm}$ = 800 GeV the performance of the cavities has to be pushed closer to the physical limit of 50-60 MV/m. This limit is determined by the properties of the superconductor niobium. However, in the present 9-cell cavities, the maximum gradient is limited at lower gradients by three main effects:
\newline
\par\textbullet \ thermal instabilities (quenches) 
\par\textbullet  \ field emission currents
\par\textbullet \ decrease of the quality factor without indication of
\par \ \ \ field emission at 25-30 MV/m.
\newline \newline The first limitation can be caused by defects in the material or weld. Therefore the sensitivity of the eddy current scanning apparatus was improved significantly for a better diagnostic of niobium sheets with tiny defects \cite{ABrinkmann}. In order to raise the onset of field emission further, efforts are undertaken to improve the high pressure rinsing system and the particle control during the assembly in the cleanroom \cite{Reschke}.  The third limitation is not well understood yet but measurements on electropolished superconducting cavities have demonstrated, that  electropolishing (EP)  can overcome this  limit. 
\subsection{Electropolished Cavities}
\begin{figure}[htb]
\centering
\includegraphics*[height=70mm, angle=-90]{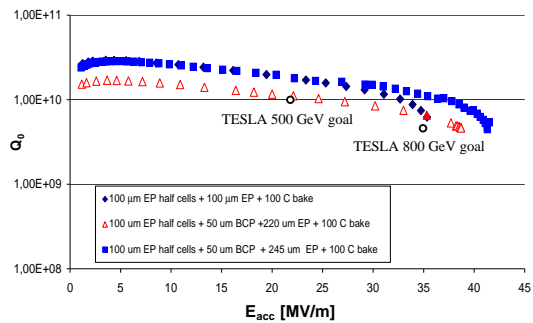}
\caption{Performance of electropolished 1-cell 1.3 GHz cavities after backout at moderate temperature ($\sim$100 $ ^{\circ}$C). For comparison also shown is the required accelerating gradient for TESLA with a centre-of-mass energy $E_{cm}$ = 500 GeV and for an  energy upgrade to $E_{cm}$ = 800 GeV.  }
\label{ep}
\end{figure}

In a collaboration between KEK and Saclay it has been shown convincingly that EP raises the maximum accelerating field in 1-cell cavities by more than 7 MV/m with respect to the standard buffered chemical polishing \cite{EPKEK}. Therefore in a joined effort between CERN, DESY, KEK and Saclay a electropolishing system for 1-cell cavities has been  set up at CERN for further investigations. Using the KEK-style electropolishing, very high gradients around 40 MV/m have been achieved in several 1-cell resonators, as shown in Fig.~\ref{ep}.
Interestingly also this cavities showed a decrease of the quality factor without indication of field emission at high gradients, before  bakeout at moderate temperature ($\sim$100 $ ^{\circ}$C). However, experiments have shown that in-situ baking at moderate temperature drops $R_{BCS}$ by a factor of 2, significantly reduces the Q-slope and improves the maximum gradients  in EP cavities \cite{Kneisel,Lutz}. Presently it is unclear, why this bakeout seems to be more efficient on EP cavities than on BCP cavities. It is remarkable, that the high performance of electropolished 1-cell cavities is  achieved without high-temperature (1400 $ ^{\circ}$C) treatment, normally done to improve the niobium quality. Presently the EP technology is transferred to the TTF 9-cell resonators in collaboration with KEK. 
\subsection{Hydroformed Cavities}
\begin{figure}[htb]
\centering
\includegraphics*[height=70mm, angle=90]{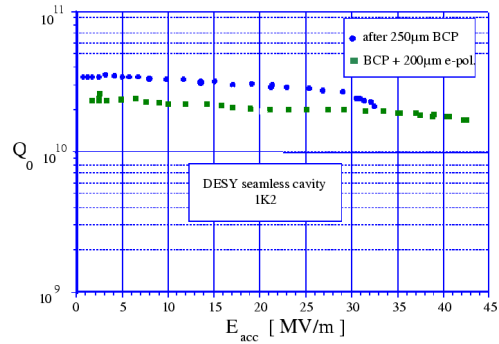}
\caption{Performance of a hydroformend 1-cell 1.3 GHz cavity before and after electorpolishing. The cavity was fabricated at DESY and tested by P. Kneisel at JLAB.}
\label{hydro}
\end{figure}

The development of seamless cavity fabrication was motivated by the potential of cost reduction, but recent tests have shown, that seamless cavities are also promising to reach highest gradients. Presently two methods of cavity fabrication without welding are studied: spinning \cite{Palmieri} and hydroforming \cite{WSinger}. The hydroforming of cavities from seamless niobium tubes is being pursued at DESY. Four 1-cell cavites have been successfully built so far. After a buffered chemical polishing, one of these resonators reached a high accelerating gradient of 32.5 MV/m at a remarkable high quality factor $Q_0=2\cdot10^{10}$. By electropolishing the inner surface of this cavity the gradient was further enhanced beyond 40 MV/m, again showing a very high quality factor, see Fig.~\ref{hydro}. 
One should note that this cavity was produced from a low $RRR$ niobium ($RRR=100$). Afer a 1400 $ ^{\circ}$C heat treatment the $RRR$ raised to 300-400.
\subsection{Superstructure}
The concept of the superstructure has been proposed for the TESLA main linac, aiming to reduce the spacing between the cavites and to save in rf components \cite{Sekuto1}. In this concept several multi-cell cavities are coupled by short beam tubes. The whole chain can be fed by one FM coupler attached at one end beam tube.  The superstructure-layout is extensively studied at DESY since 1999. Computations have been performed for the rf properties of the cavity-chain, the bunch-to-bunch energy spread and multibunch dynamics. A copper model of a 4 x 7-cell superstructure has been built in order to compare with the simulations and for testing the field-profile tuning and the HOM damping scheme \cite{Sekuto2}. A "proof of principle" niobium prototype of the 4 x 7-cell superstructure is now under construction and will be tested with beam at the TESLA Test Facility in 2001 \cite{Sekuto3}.
\section{PULSED OPERATION OF SUPERCONDUCTING CAVITIES}
\begin{figure}[htb]
\centering
\includegraphics*[width=65mm, height=40mm]{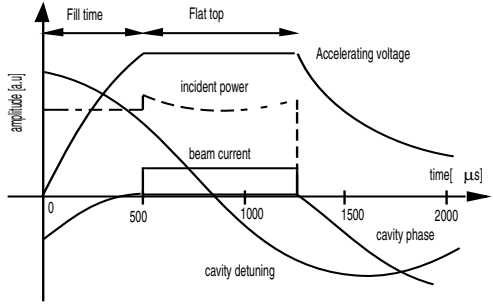}
\caption{Schematic view of the pulse structure of the TTF cavity operation. The accelerating field is increased during  0.5 ms filling time, followed by 0.8 ms constant gradient.}
\label{pulse}
\end{figure}

The superconducting cavities at the TTF-linac are operated in pulsed mode. The pulse structure of the rf power and the beam is shown in Fig.~\ref{pulse}.
The rf pulse length is 1300 $\mu$s from which 500 $\mu$s are required to fill the cavity. Amplitude and phase control is obviously needed during the flat-top of 800 $\mu$s when beam is accelerated, but it is equally desirable to control the field during filling to ensure proper beam injection conditions. The pulsed structure of the rf field and the beam current puts demanding requirements on the rf control system. There are two major sources of  field perturbations in a superconducting cavity: 
\newline \par
\par\textbullet\  beam loading and fluctuation of the beam current
\par\textbullet  \ modulation of the cavity resonance frequency.
\newline \newline The resonance frequency is modulated by  deformations of the cavity walls induced by  external mechanical vibrations (microphonics) or by the time dependent Lorentz force, see Fig.~\ref{pulse}. The resulting amplitude and phase perturbations are in the order of 5$\%$ and $20^{\circ}$, respectively. These errors must be suppressed by one to two orders of magnitude by the rf control system. Fortunately, the dominating  perturbations (Lorentz force detuning and beam loading) are highly repetitive from pulse to pulse. Therefore they can be reduced by use of feedforward compensation significantly. The  feedforward control is optimized continously by making the feedforward system adaptive \cite{Liepe}. The remaining non repetitive errors can be suppressed by feedback control. 
\subsection{RF Field Control}
\begin{figure}[htb]
\centering
\includegraphics*[width=70mm]{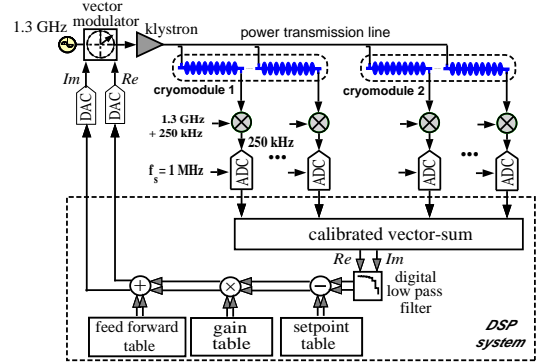}
\caption{Schematic of the digital rf control system.}
\label{control}
\end{figure}

The two modules installed in the TTF-linac are supplied with rf power by one 10 MW multi-beam klystron. Amplitude and phase control can be accomplished by modulation of the incident rf wave which is common to the 16 cavities. Therefore control of an individual cavity field is not possible, but the  sum of the electric field vectors of all 16 cavities is regulated by a digital rf control system \cite{Schilcher}. The layout of the TTF digital rf control system is based on controlling the real and imaginary part of the complex vector sum, as shown in Fig.~\ref{control}. 
The  rf control system is realized as a combination of feedback and feedforward control. The effectiveness of the feedback system is limited by the loop delay of 5 $\mu$s and the unity-gain bandwidth of about 20 kHz. When feedback is applied, the residual control error  is dominated by a repetitive component. At the TTF linac it could be demonstrated, that this error can be reduced further by more than an order of magnitude with the adaptive feedforward \cite{Gamp}. The  high degree of field stability achieved at the TTF linac is mainly due to the low level of fast fluctuations, like microphonics. 
\subsection{Microphonics}
External mechanical vibrations can change the shape of the cavity and thereby the eigenfrequency of the cavity. This results in an amplitude and phase  jitter of the accelerating field. Superconducting cavities are sensitive to microphonics due to the thin wall thickness and the small bandwidth. The rms microphonics frequency spread, measured in 16 cavities of the TTF-linac, is 9.5$\pm$5.3 Hz and thus surprisingly small for a superconducting cavity system. The spectrum of this frequency fluctuations   is dominated be a few frequencies. 
\subsection{Lorentz-Force Detuning}
\begin{figure}[htb]
\centering
\includegraphics*[width=65mm]{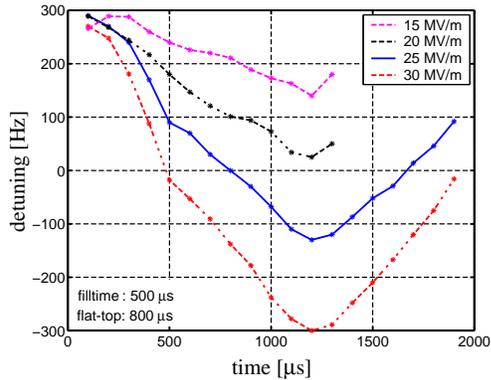}
\caption{Dynamical detuning of a TTF 9-cell cavity during pulsed operation at different flat-top gradients.}
\label{detuning}
\end{figure}

The resonance frequency of pulsed cavities is modulated by the time-varying Lorentz force within the rf pulse, see Fig.~\ref{pulse}. The dynamic Lorentz force detuning is correlated with the pulsed rf field and is depending on the mechanical properties of the cavity. The steady state Lorentz force detuning at constant accelerating gradient $E_{acc}$ is proportional to the square of the gradient: $\Delta f = -K\cdot E^2_{acc}$. The TTF cavities have been designed for a steady state detuning constant $K=$ 1 Hz/(MV/m)$^2$. This quadratical dependence is also reflected in the dynamic Lorentz force detuning during the rf pulse. Figure ~\ref{detuning} shows the fast inrease of the dynamic frequency shift at high flat-top gradients.   
Preliminary measurements at CHECHIA and the TTF-linac indicate, that the dynamic detuning during the flat-top ranges form $\pm$120 Hz (total drift 240 Hz) to $\pm$300 Hz  at the design gradient for TESLA500. The corresponding additional rf power required to maintain a constant accelerating field during the flat-top amounts to $\sim$5$\%$ to $\sim$30$\%$. Given that the exta rf power for cavity field control should not exceed 10$\%$ this means that the stiffening of the cavities for TESLA must be better controlled or increased. 
\subsection{Active Compensation of Lorentz-Force Detuning}
\begin{figure}[htb]
\centering
\includegraphics*[width=65mm]{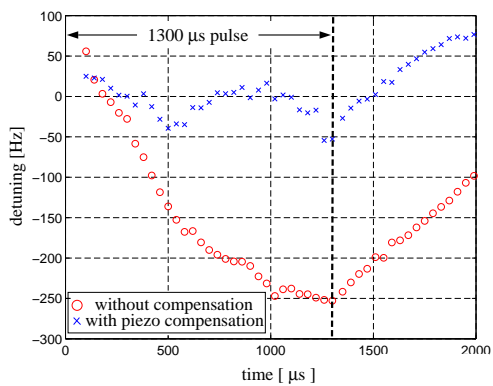}
\caption{Active compensation of Lorentz-force detuning during the rf pulse. In the first successfull test show here, a TTF 9-cell cavity is operated at 20 MV/m flat-top gradient.}
\label{comp}
\end{figure}

In case that the present passive stiffening of the TTF cavities does not limit the additional power to 10$\%$, it will be necessary to increase the stiffening (e.g. by copper spraying \cite{spraying}) or to implement a fast active frequency control scheme. A solution for a fast frequency tuner could be based on piezotranslators. The piezotranslator would allow for a fast frequency tuning within the rf pulse to compensate the detuning induced by the Lorentz force. A proof of principle experiment of this compensation with a fast piezotuner has been conducted successfully at the horizontal test stand at DESY \cite{piezo}, see Fig.~\ref{comp}.  Such a system appears especially attractive  since it will improve the overall efficiency of the TESLA linacs singnificantly. More detaild studies will be done in the near future to determine the performance limitations of a fast piezotuner.
\section{CONCLUSIONS}
The presented results  demonstrate the availability of the superconducting technology for the 500 GeV linear collider TESLA. The TTF 9-cell cavities are now routinely reaching the TESLA requirements of 22 MV/m.  The third module in the TTF-linac was successfully operated  at the TESLA design gradient. The technology needed for the pulsed operation at high gradient was developed and the reliable operation  of a superconducting linac  in pulsed mode was demonstrated at the TTF-linac. Recent results of 1-cell cavities justify the optimism that the accelerating gradient in the 9-cell cavities  can be increased even further.   

\end{document}